\documentclass[12pt]{article}
\usepackage{a4wide}

\usepackage[dvipdfmx]{graphicx} 
\usepackage{mediabb}                    

\usepackage{color}

\newcommand{\al}[1]{\begin{align}#1\end{align}}

\newcommand{\paren}[1]{\left(#1\right)}

\newcommand{\sqbr}[1]{\left[#1\right]}
\newcommand{\br}[1]{\left\{#1\right\}}

\newcommand{\nn}{\nonumber\\}

\newcommand{\p}{\partial}

\usepackage{amsmath,amssymb}
\usepackage{epsf}

\begin{document}
\title{\vbox{
\baselineskip 14pt
\hfill \hbox{\normalsize MAD-TH-17-01}
} \vskip 1cm
\bf \Large 
Memory in de Sitter space and 
\\
BMS-like supertranslations
\vskip 0.5cm
}
\author{
Yuta~Hamada$^{1,2}$,
Min-Seok~Seo$^{3}$,
Gary~Shiu$^{1}$,
\bigskip\\
$^1$\it\normalsize
Department of Physics, University of Wisconsin-Madison, Madison, Wisconsin 53706, USA\\
$^2$\it\normalsize
KEK Theory Center, IPNS, KEK, Tsukuba, Ibaraki 305-0801, Japan\\
$^3$\it\normalsize
 Center for Theoretical Physics of the Universe,\\
\it\normalsize Institute for Basic Science (IBS), Daejeon 34051, Korea
}
\date{\today}


\maketitle

\abstract{\noindent \normalsize
It is well known that the memory effect in flat spacetime is parametrized by the BMS supertranslation.
We investigate the relation between the memory effect and diffeomorphism in de Sitter spacetime.
We find that gravitational memory is parametrized by a BMS-like supertranslation in the static patch of de Sitter spacetime.
We also show a diffeomorphism that corresponds to gravitational memory in the Poincare/cosmological patch.
Our method does not need to assume the separation between the source and the detector to be small compared with the Hubble radius,
and can potentially be applicable to other FLRW universes, as well as ``ordinary memory" mediated by massive messenger particles.}

\newpage
\normalsize
\section{Introduction}
Understanding the vacuum structure of gravity has been a long-standing line of inquiry in general relativity. 
It was recently reemphasized that, even in flat spacetime, 
the vacuum for Einstein gravity is infinitely degenerate~\cite{Strominger:2013jfa}.
This degeneracy can be physically observable in terms of the so called memory effect, which can be understood as a transition between vacua~\cite{Strominger:2014pwa}.
In brief, the memory effect is a permanent change in the relative separation of test particles that make up the detector induced by the passage of a pulse. This permanent effect induced by a gravitational wave pulse has come to be known as the gravitational memory.
The vacuum transition that induces the memory
is expected to be generated by spontaneously broken spacetime symmetries, and thus should be described
by a subgroup of diffeomorphism.
Interestingly, in the case of flat spacetime, the memory effect
was shown~\cite{Strominger:2014pwa,Hollands:2016oma} to be
completely described in terms of BMS supertranslation, the asymptotic symmetries of flat spacetime uncovered in the 1960s by Bondi, Metzner, van der Burg and Sachs~\cite{Bondi:1962px,Sachs:1962wk,Sachs:1962zza}.

The resurgence of interest in the Bondi-Metzner-Sachs (BMS) supertranslation symmetries is due largely to their possible role in the black hole information paradox. The existence of infinitely many vacua might be attributed to a new kind of soft black hole hair~\cite{Dvali:2015rea,Hawking:2016msc,Averin:2016ybl,Hawking:2016sgy,Averin:2016hhm}.
Although the no hair theorem states that any static black hole in the Einstein-Maxwell theory can be completely characterized by its mass, charge, and angular momentum (up to diffeomorphism), the BMS transformation changes the physical state, and therefore,  it is logically possible for additional information to be attributed to the BMS charges.
While the role of the BMS charges in the black hole information paradox is still a subject of active discussions (see e.g., \cite{Mirbabayi:2016axw}), we shall focus on the relation between diffeomorphism and the memory effect which is currently better established.

In fact, the memory effect is part of a ``triangular relation" \cite{Strominger:2014pwa, He:2014laa} not only with BMS symmetries but also with soft graviton theorems.
Weinberg's soft photon/graviton theorems relate scattering 
amplitudes with insertions of soft photon/gravitons to that without the insertions.
These soft theorems were shown to follow from the Ward identities of BMS supertranslation~\cite{He:2014laa}.
The aforementioned `triangular relation'
may thus shed light on the infrared structure of quantum gravity.
As these infrared properties are dictated by symmetries, they should not depend on the specific form of quantum gravity.

Aside from these theoretical considerations, memory effect is also observationally 
interesting as it probes the infrared properties of gravity. Gravitational memories generated by astrophysical events are potentially detectable in future gravity wave experiments such as LISA, which has sensitivity to the low frequency bands. Gravity wave interferometers with longer arm length designs, such as the proposed Taiji experiment \cite{Taiji} may even have better sensitivity. 
In light of these experimental prospects, it is natural to ask whether the above triangular relation applies to an accelerating universe like ours as well.
The effect of gravitational memory in Minkowski spacetime was derived long ago~\cite{Zeldovich1974,Braginsky:1986ia,Braginsky1987}, and has since been generalized to de Sitter spacetime or FLRW universes by several groups \cite{Chu:2015yua,Bieri:2015jwa,Kehagias:2016zry,Chu:2016qxp,Tolish:2016ggo,Chu:2016ngc}.
 Our derivation of the memory effect in de Sitter space differs from 
 that of previous works in several respects, most notably, the emphasis on its relation
 with spacetime diffeomorphisms. Our approach parallels to that carried out for flat spacetime \cite{Strominger:2014pwa}, but without the restrictions on the spacetime asymptotics, and thus we believe it has wider applicability to other cosmological contexts.
We identify a subgroup of diffeomorphism of de Sitter spacetime which corresponds to gravitational memory, which as we will see is different from the asymptotic symmetries of de Sitter spacetime \cite{Anninos:2010zf,Ashtekar:2014zfa,Koga:2001vq}.
We find that, in the static patch of de Sitter spacetime, a BMS-like supertranslation is equivalent to the memory effect, as in the case of flat spacetime.
We also find a diffeomorphism that corresponds to the memory effect in the Poincare/cosmological patch.

It is instructive to compare our approach with previous investigations on the memory effect in de Sitter spacetime~\cite{Chu:2015yua,Bieri:2015jwa,Chu:2016qxp,Tolish:2016ggo,Chu:2016ngc} .
Our definition of the memory effect is different from that in Refs~\cite{Chu:2015yua,Chu:2016qxp,Chu:2016ngc}, as the effect of tidal force is naturally excluded in our analysis. In contrast to Ref.~\cite{Bieri:2015jwa}, 
our method does not need to assume a small Hubble scale in comparison to the separation between the source and the detector. 
As compared to Ref.~\cite{Tolish:2016ggo}, we do not rely on linearized perturbation theory, and extended sources/detectors can be treated.
While we have only carried out our study for null memory in de Sitter spacetime, our approach is potentially applicable to other FLRW 
universes, and memory effect due to massive messengers (known as ``ordinary memory").

This paper is organized as follows. In Sec.~\ref{Sec:memory}, we introduce the setup of the memory effect.
In Sec.~\ref{Sec:BMS}, we review the BMS transformation in asymptotic flat spacetime, and identify the so-called BMS supertranslation and superrotation. In Sec.~\ref{Sec:flat memory}, we show the relation between memory effect and BMS supertranlsation in flat spacetime. Our presentation is particularly suited for uncovering a similar relation to de Sitter space which we present in Sec.~\ref{Sec:dS memory}. Sec.~\ref{Sec:Summary} is devoted to a summary and conclusions. 

\section{The memory effect}\label{Sec:memory}
\begin{figure}[t]
\begin{center}
\hfill
\includegraphics[width=.7\textwidth]{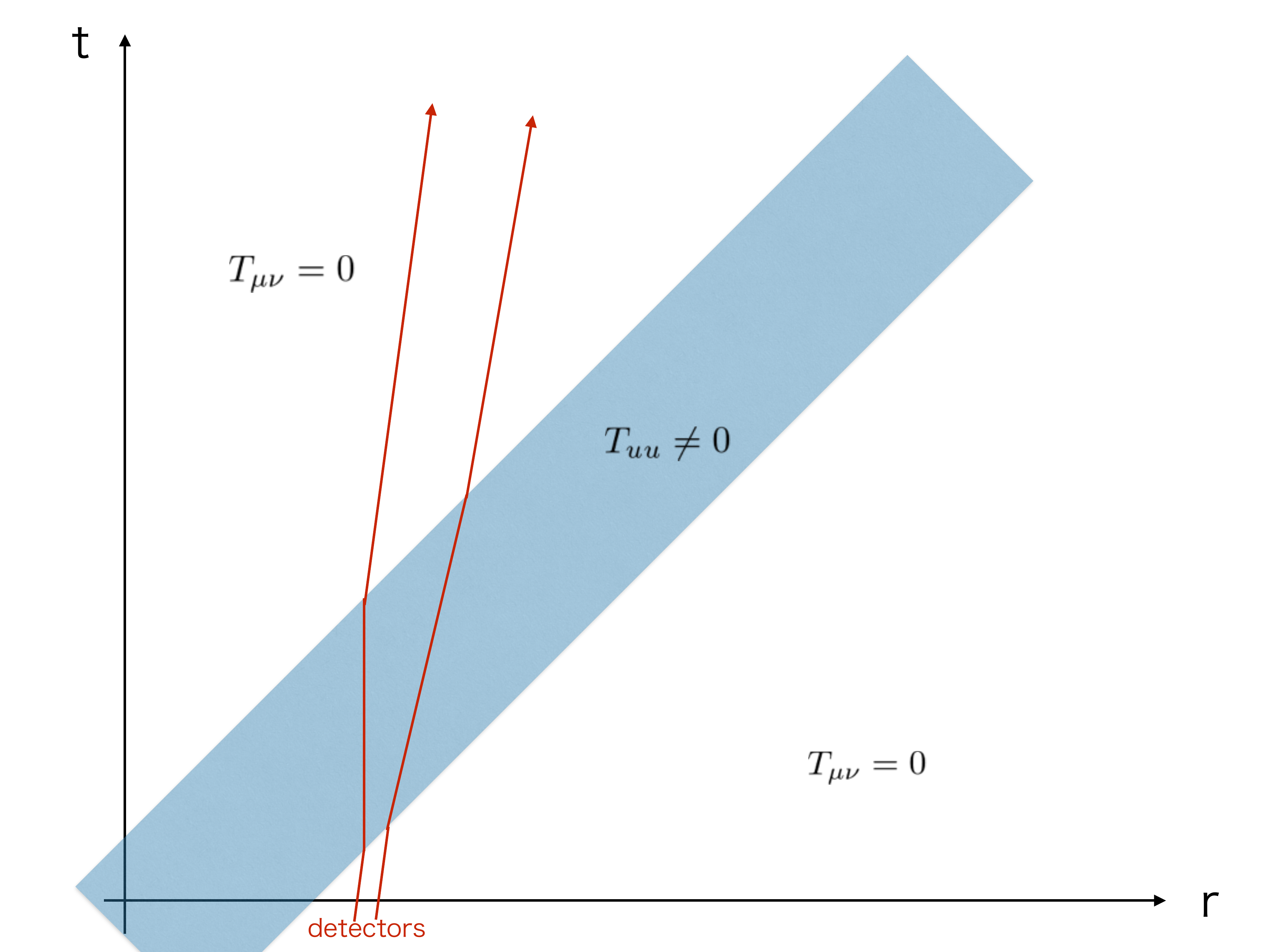}
\hfill\hfill\mbox{}
\end{center}
\caption{
A schematic picture of the memory effect.
The red line corresponds to the worldline of the detector.
The blue region represents the emission of a pulse, $T_{uu}\neq0$.
Except for the blue region, the energy momentum tensor is zero.
The memory effect is a change of the distance of the detector before and after the emission of a pulse.
}
\label{Fig:memory}
\end{figure}
Let us consider two 
detectors with a non-zero distance between them.
The memory effect is the change of the distance of the detectors, $L$, before and after a pulse injection. 
This effect is known as gravitational memory if the pulse is a gravitational wave, whereas an electromagnetic pulse gives rise to electromagnetic memory.
In terms of the retarded time $u=t-r$, the memory effect which we consider in this paper can be described as follows:
\begin{enumerate}
\item $u<u_i$

The detector is in the vacuum, namely, $T_{\mu\nu}=0$.
The corresponding metric $g_{\mu\nu}^{\text{ini}}$ is a solution of the vacuum Einstein equation.
The distance between the two detectors is given by $L=\int\sqrt{g_{\mu\nu}^{\text{ini}}dx^\mu dx^\nu}$.

\item $u_i<u<u_f$

The $uu$ component of the energy  momentum tensor is not zero, $T_{uu}\propto 1/r^2$, due to the pulse injection. Notice that, in a general FLRW background, the conservation of energy momentum tensor, $\nabla_\mu T^{\mu\nu}=0$, is consistent with $T_{uu}\neq0$ and vanishing other components if $T_{uu}\propto r^{-2}$, where $r$ is the proper distance.
The metric evolves following the Einstein equation.

\item $u_f<u$

The detector is again in the vacuum, $g_{\mu\nu}^{\text{fin}}$.
However, in general, $g_{\mu\nu}^{\text{fin}}\neq g_{\mu\nu}^{\text{ini}}$, and the difference of the distance $\Delta L=\int\sqrt{g_{\mu\nu}^{\text{fin}}dx^\mu dx^\nu}-\int\sqrt{g_{\mu\nu}^{\text{ini}}dx^\mu dx^\nu}$ is called the memory effect.
\end{enumerate}
The schematic picture is shown in Fig.~\ref{Fig:memory}.
The red line corresponds to the worldlines of the two detectors.
In the blue region, the energy is injected, and the $uu$ component of the energy momentum tensor is nonzero. 

It would be difficult to detect this effect with ground-based detectors such as LIGO because it is a low frequency effect.
Detectors which are sensitive to the low frequency band like LISA (and even longer arm length designs such as the proposed Taiji experiment \cite{Taiji})
might have the detection possibility. 
See e.g. Ref.~\cite{Favata:2010zu} for a realistic estimation of the detectability.

\section{BMS transformation}\label{Sec:BMS}
In this section, we briefly review the asymptotic symmetry of asymptotically flat spacetime, which was originally considered by Bondi, Metzner, van der Burg and Sachs in the 1960s~\cite{Bondi:1962px,Sachs:1962wk,Sachs:1962zza}.

Throughout the section, we utilize the outgoing Bondi coordinate $(u,r,z,\bar{z})$, and employ the Bondi gauge, where
\al{\label{Eq:Bondi gauge}
&
g_{rA}=g_{rr}=0,
&&
{\rm det}\paren{{g_{AB}\over r^2}}
=
\paren{\gamma_{z\bar{z}}}^2.}
Here $A, B$ represent $z$ and $\bar{z}$, and $\gamma_{z\bar{z}}=2/(1+|z|^2)^2$.
The $z$ coordinate is related to the standard polar coordinate by $z=e^{i\phi}\tan\theta/2$.
The asymptotic flat space is defined as the spacetime which becomes close to the flat one around null infinity with an appropriate fall-off function.
Then, the asymptotic symmetry is defined as the subgroup of diffeomorphism which does not modify the gauge and the fall-off conditions.

More concretely, the asymptotic form of the metric is
\al{
ds^2&=-du^2-2dudr
+2r^2\gamma_{z\bar{z}}dzd\bar{z}
\nn&+\mathcal{O}\paren{1\over r}du^2+\mathcal{O}\paren{1\over r^2}dudr+\mathcal{O}\paren{1}dudz+\mathcal{O}\paren{1}dud\bar{z}
\nn&+\mathcal{O}\paren{r}dz^2+\mathcal{O}\paren{r}d\bar{z}^2+\mathcal{O}\paren{r}dzd\bar{z},
}
and the asymptotic symmetry should satisfy~\cite{Sachs:1962zza}
\al{&
\delta g_{rr}=\delta g_{rz}=g^{AB}\delta g_{AB}=0,
\nn
&
\delta g_{uu}=\mathcal{O}\paren{1\over r},
\quad
\delta g_{uz}=\mathcal{O}\paren{1},
\quad
\delta g_{ur}=\mathcal{O}\paren{1\over r^2},
\quad
\delta g_{AB}=\mathcal{O}\paren{r},
}
where $\delta g_{\mu\nu}$ is the change of the metric corresponding to the infinitesimal diffeomorphism.
Let us examine the solution of the above conditions. 
The general transformation is 
\al{
&
u\to u+\epsilon_u(u,r,z,\bar{z}),
&&
r\to r+\epsilon_r(u,r,z,\bar{z}),
\nn
&
z\to z+\epsilon_z(u,r,z,\bar{z}),
&&
\bar{z}\to \bar{z}+\epsilon_{\bar{z}}(u,r,z,\bar{z}),
}
The condition $\delta g_{rr}=0$ just implies that $\epsilon_u(u,r,z,\bar{z})=
\epsilon_u(u,z,\bar{z})$.
Then, $\delta g_{rz}=0$ reads
\al{&
\epsilon_z=
-{1\over r}\partial^{z}\epsilon_u
+
a_z(u,z,\bar{z})
+
\mathcal{O}\paren{1\over r^2}
,
&&
\epsilon_{\bar{z}}=
-{1\over r}\partial^{\bar{z}}\epsilon_u
+
a_{\bar{z}}(u,z,\bar{z})
+
\mathcal{O}\paren{1\over r^2}
.
}
The last condition $g^{AB}\delta g_{AB}=0$ requires $\delta g_{z\bar{z}}=\mathcal{O}(r^{-2})$, which is equivalent to
\al{
\epsilon_r
=
D^zD_z\epsilon_u-r
\br{
\partial^{\bar{z}}\paren{{a_z\over\paren{1+|z|^2}^2}}
+
\partial^z\paren{{a_{\bar{z}}\over\paren{1+|z|^2}^2}}
}
   +\mathcal{O}(r^{-1})
   .
   }
This is the general transformation which is compatible with gauge fixing.
Let us consider the consequences of the fall-off conditions.
The $\delta g_{zz}, \delta g_{\bar{z}\bar{z}}=\mathcal{O}(r)$ and $\delta g_{uz}=\mathcal{O}(1)$ conditions give
\al{&\label{Eq:az condition}
\partial_{\bar{z}}a_z=\partial_{z}a_{\bar{z}}=0,
&&\partial_u a_z=\partial_u a_{\bar{z}}=0.
}
The remaining conditions are satisfied if
\al{
\epsilon_u=
-f(z,\bar{z})
+
\int du
\sqbr{
\partial^{\bar{z}}\paren{{a_z\over\paren{1+|z|^2}^2}}
+
\partial^z\paren{{a_{\bar{z}}\over\paren{1+|z|^2}^2}}
}
+\mathcal{O}\paren{1\over r}
.
}
To summarize, the residual gauge transformation is
\al{\label{Eq:BMS transformation}
&
\epsilon_u=
-
f(z,\bar{z})
+
\int du
\sqbr{
\partial^{\bar{z}}\paren{{a_z(z)\over\paren{1+|z|^2}^2}}
+
\partial^z\paren{{a_{\bar{z}}(\bar{z})\over\paren{1+|z|^2}^2}}
}
,
\nn
&
\epsilon_r=D^zD_z\epsilon_u
-r
\sqbr{
\partial^{\bar{z}}\paren{{a_z\over\paren{1+|z|^2}^2}}
+
\partial^z\paren{{a_{\bar{z}}\over\paren{1+|z|^2}^2}}
}
,
\nn
&
\epsilon_z=
-{1\over r}\partial^{z}\epsilon_u
+
a_z(z)
,
\nn
&
\epsilon_{\bar{z}}=
-{1\over r}\partial^{\bar{z}}\epsilon_u
+
a_{\bar{z}}(\bar{z})
.
}
One can see that the asymptotic transformation consists of a real function $f(z,\bar{z})$ and a holomorphic function $a_z(z)$.
The former and the latter are known as BMS supertranslation~\cite{Bondi:1962px,Sachs:1962wk,Sachs:1962zza} and superrotation~\cite{Barnich:2009se,Barnich:2011mi}, respectively.
These are generalization of global translation and rotation.
In fact, 
\al{\label{Eq:global translation}
f=c_0+{c_1\paren{1-|z|^2}+c_2\bar{z}+\bar{c}_2z\over 1+|z|^2},
}
corresponds to translation of time ($c_0$) and spatial coordinates ($\mathrm{Re}(c_2), \mathrm{Im}(c_2), c_1$), and
\al{
a_z=\epsilon_1 +\epsilon_z z+\epsilon_{z2} z^2
}
corresponds to boost and spatial rotation.

\section{Memory in Minkowski spacetime}\label{Sec:flat memory}
In this section, the relation between memory and BMS supertranslations is reviewed.
As we have seen in the previous sections, there exists a surprisingly close relationship between these two seemingly different notions~\cite{Strominger:2014pwa}.

Since we are interested in the region where $r$ is much larger than the typical size of the source,  we can consider a
 large $r$ expansion, and the metric becomes\footnote{We have omitted $\ln r$ terms for simplicity~\cite{Barnich:2011mi}, which do not affect our discussion.}~\cite{Barnich:2011mi,Hawking:2016sgy}:
\al{\label{Eq:flat}
ds^2&=
-du^2-2dudr+2r^2\gamma_{z\bar{z}}dzd\bar{z}
\nn
&+{2m_B(u,z,\bar{z})\over r}du^2+rC_{zz}(u,z,\bar{z})dz^2+rC_{\bar{z}\bar{z}}(u,z,\bar{z})d\bar{z}^2\nn
&+U_z(u,z,\bar{z})dudz+U_{\bar{z}}(u,z,\bar{z})dud\bar{z}
\nn
&+{2m_B^{(2)}(u,z,\bar{z})\over r^2}du^2+{1\over r^2}D_{ur}(u,z,\bar{z})dudr
\nn&+{1\over r}\paren{{4\over3}N_z(u,z,\bar{z})+{4\over3}u\partial_z m_B-{1\over4}\partial_z(C_{zz}C^{zz})}dudz
\nn&+{\paren{1+|z|^2}^2\over4}C_{zz}C_{\bar{z}\bar{z}}dzd\bar{z}+....
}
Here the indices $z,\bar{z}$ are raised and lowered by $\gamma_{z\bar{z}}$, $N_z$ is related to the angular momentum aspect, and the last term is fixed by the requirement of the Bondi gauge.
We solve the Einstein equation, which relates the above variables, at $\mathcal{O}(r^{-3})$, $\mathcal{O}(r^{-2})$, $\mathcal{O}(r^{-1})$ for  the $ab$, $aA$ and $aA$ components, where $a=u,r $ and $A=z, \bar{z}$.
From the $rz$ component at $\mathcal{O}(r^{-1})$, we obtain $U_z=D^z C_{zz}$, where $D_z$ is the $\gamma_{z\bar{z}}$-covariant derivative.
The relation 
\al{
D_{ur}={(1+|z|^2)^4\over16}C_{zz}C_{\bar{z}\bar{z}}+h(z,\bar{z})
}
can be obtained from the $z\bar{z}$ component.
Here, $h(z,\bar{z})$ is an arbitrary function of $(z,\bar{z})$.
The $uu$ component of the Einstein equation imposes
\al{\label{Eq:uu relation}
&\partial_u m_B&={1\over4}\sqbr{D_z^2N^{zz}+D_{\bar{z}}^2N^{\bar{z}\bar{z}}}-T_{uu},
&&T_{uu}={1\over4}N_{zz} N^{zz}+4\pi \lim_{r\to\infty}r^2T_{uu}^{\text{(matter)}}
}
at $\mathcal{O}(r^{-2})$. 
Here $N_{zz}=\partial_u C_{zz}$ is the Bondi news.
Similarly,  the $\mathcal{O}(r^{-3})$ relation determines the $u$ dependence of $m_B^{(2)}$. 
The $uz$ component yields
\al{\label{Eq:uz relation}
\partial_u N_z&=
-{1\over4}D^{\bar{z}}\gamma^{z\bar{z}}\paren{D_{\bar{z}}^2C_{zz}-D_z^2C_{\bar{z}\bar{z}}}+u\partial_z\paren{T_{uu}-{1\over4}\paren D_z^2N^{zz}+D_{\bar{z}}^2N^{\bar{z}\bar{z}}}-T_{uz},
\nn
T_{uz}&=
8\pi\lim_{r\to\infty}r^2T_{uA}^{\text{(matter)}}-{1\over 4}\partial_z\paren{C_{zz}N^{zz}}
-{1\over2}C^{\bar{z}\bar{z}}\partial_z N_{\bar{z}\bar{z}}
}
at $\mathcal{O}(r^{-2})$, where we have used Eq.~\eqref{Eq:uu relation}.
In the vacuum, $N_{zz}=0, T_{uA}^{\text{(matter)}}=0$ and $\partial_u N_z=0$ are satisfied, and Eq.~\eqref{Eq:uz relation} becomes $D_z\paren{D_{\bar{z}}^2C_{zz}-D_{z}^2C_{\bar{z}\bar{z}}}=0$.
Then, requiring that the metric to be globally defined on $S^2$, the general solution of the last condition is
\al{\label{Eq: flat memory}
C_{zz}=2D_z^2 f(z,\bar{z}),
}
where $f$ is a real function, and can be expressed as a superposition of the spherical harmonics.
Obviously, this is the same as the transformation law of BMS supertranslations with the parameter $f$.

Without loss of generality, we take $f=0$ at the initial state $u<u_i$, and denote the vacuum at $u_f<u$ by $f=f_\text{fin}$.
Let us consider the case where the two detectors are placed with a displacement $\delta z$.
The proper distance between the  detectors is given by\footnote{Since we are considering a large $r$ expansion, $r^{-1}$ correction always exists.} 
\al{
L\simeq{2r|\delta z|\over 1+|z|^2},
}
for $u<u_i$, and after $u=u_f$ the change of the distance $\Delta L$ is
\al{
\Delta L\simeq{r\over2L}\Delta C_{zz} \paren{\delta z}^2+c.c..
}
Therefore, the memory effect is nothing but the change of $C_{zz}$. In this sense, we see from Eq.~\eqref{Eq: flat memory} that the memory effect  is parametrized by the BMS supertranslation.

The explicit relation between memory and the supertranslation parameter $f$ is given by solving Eq.~\eqref{Eq:uu relation}.
Integrating over $u$, we obtain
\al{\label{Eq:u integral}
m_B\big|^{u=u_f}_{u=u_i}&=
{1\over4}\sqbr{D_z^2C^{zz}+D_{\bar{z}}^2C^{\bar{z}\bar{z}}}\bigg|^{u=u_f}_{u=u_i}-\int^{u_f}_{u_i}du T_{uu},
\nn&=(D^z)^2D_z^2f_\text{fin}-\int^{u_f}_{u_i}du T_{uu},
}
from which we have\footnote{
Here we define the measure as $\gamma_{z\bar{z}}d^2z:=\gamma_{z\bar{z}}d\paren{\tan^2{\theta\over2}}d\phi=\sin\theta d\theta d\phi$, and the delta function is defined as $\int d^2z \delta^2(z-z'):=1$.}
~\cite{Strominger:2014pwa}
\al{\label{Eq:f solution}
f_\text{fin}(z,\bar{z})&= -\int d^2z' \gamma_{z'\bar{z}'}G(z,\bar{z};z',\bar{z}')\paren{m_B\big|^{u=u_f}_{u=u_i}+\int^{u_f}_{u_i}du T_{uu}},
\nn
G(z,\bar{z};z',\bar{z}')&:=-{1\over\pi}\sin^2{\Theta\over2}\log\sin^2{\Theta\over2},
\quad
\sin^2{\Theta\over2}={|z-z'|^2\over(1+z'\bar{z}')(1+|z|^2)}.
}
In fact, $D_{\bar{z}}^2 D_{z}^2 G$ gives,
\al{\label{Eq:Green function}
&D_{\bar{z}}^2 D_{z}^2 G(z,\bar{z};z',\bar{z}')
\nn
&=
-\gamma_{z\bar{z}}\delta^2\paren{z-z'}
+{1\over\pi^2(1+|z|^2)^5}\sqbr{\paren{1+|z|^2}+{3\paren{1-|z|^2}\paren{1-|z'|^2}+6z\bar{z}'+6\bar{z}z'\over1+|z'|^2}},
}
where $\partial_{\bar{z}}z^{-1}=2\pi\delta^2(z)$ is used.
The integration of first term
with respect to $z'$ reproduces Eq.~\eqref{Eq:u integral}.
On the other hand, the second and third terms correspond to the $l=0, 1$ components of the spherical harmonics of $z'$, respectively. 
By using the orthogonal relation of the spherical harmonics $Y_{lm}$,
\al{
\int d^2z \,\gamma_{z\bar{z}}Y_{lm} Y_{l'm'}^*\propto\delta_{ll'}\delta_{mm'}.
}
one can show that the combination $m_B\big|^{u=u_f}_{u=u_i}+\int^{u_f}_{u_i}du T_{uu}$ does not contain the $l=0, 1$ components because
\al{
\int d^2z \,\gamma_{z\bar{z}}Y_{lm}\paren{m_B\big|^{u=u_f}_{u=u_i}+\int^{u_f}_{u_i}du T_{uu}}
&=\int d^2z \,\gamma_{z\bar{z}}Y_{lm}(D^z)^2D_z^2f_\text{fin}
\nn
&=\int d^2z\,\gamma_{z\bar{z}}  \paren{(D^z)^2Y_{lm}}D_z^2f_\text{fin}
\nn
&=0
}
for $l=0$ and $1$.
As a result, the integration in Eq.~\eqref{Eq:f solution} corresponding to the second and third terms of Eq.~\eqref{Eq:Green function} vanishes.

The solution of Eq.~\eqref{Eq:u integral} has an ambiguity corresponding to the zero mode of $(D^z)^2D_z^2$.
When the global finiteness on $S^2$ is required, such a zero mode is written as
\al{\label{Eq:zero mode}
f_\text{zero}
=
&c_0+{{c_1\paren{1-|z|^2}+ c_2\bar{z}+\bar{c}_2z}\over 1+|z|^2},
}
which corresponds to the usual spacetime translation (Eq.~\eqref{Eq:global translation}), and the ambiguity does not affect the memory effect.

\section{Memory in de Sitter spacetime}\label{Sec:dS memory}
Let us turn to the de Sitter universe.
As in the previous section, we show that the BMS-like supertranslation parametrizes the infinitely degenerated vacua.
In the case of de Sitter spacetime, as pointed out in Ref.~\cite{Bieri:2015jwa}, not all gravitation radiation can reach the null infinity since it is space-like. For this reason, the de Sitter null infinity is not appropriate for investigating the memory effect. 
Therefore, we consider the situation where the detector is placed near the light-like surface (cosmological horizon) with sufficiently large $r$ and employ the Bondi gauge as an appropriate description. 
We also assume that the typical frequency from the source is much larger than the Hubble scale $H$, and neglect the ``tail" effect~\cite{Chu:2015yua,Chu:2016qxp} in the following discussion.

\subsection{Static patch}\label{Sec:static patch}
We consider the de Sitter spacetime perturbed by a source at the origin.
In the large $r, H^{-1}$ expansion, we obtain
\al{\label{Eq:static patch}
ds^2&=
(-1+r^2H^2)du^2-2dudr+2r^2\gamma_{z\bar{z}}dzd\bar{z}
\nn&+\br{{m_B(u,z,\bar{z})\over r}-rH^2h_{uu}(u,z,\bar{z})}du^2+rC_{zz}(u,z,\bar{z})dz^2+rC_{\bar{z}\bar{z}}(u,z,\bar{z})d\bar{z}^2
\nn&+\br{U_{z1}(u,z,\bar{z})+H^2r^2U_{z2}(u,z,\bar{z})}dudz
\nn&+\br{U_{\bar{z}1}(u,z,\bar{z})+H^2r^2U_{\bar{z}2}(u,z,\bar{z})}dud\bar{z}+{1\over r^2}\paren{D_{ur1}+H^2r^2D_{ur2}}dudr
\nn&+{1\over r}\paren{{4\over3}N_z+{4\over3}u\partial_z m_B-{1\over4}\partial_z(C_{zz}C^{zz})}dudz+{\paren{1+|z|^2}^2\over4}C_{zz}C_{\bar{z}\bar{z}}dzd\bar{z}+....
}
The first line corresponds to the de Sitter background, and the remaining terms are perturbations.
Note that $r$ is the proper distance, and $u=t-r_*$ where $r_*$ is the tortoise coordinate.
We can define the BMS-like supertranslation in the static patch of de Sitter spacetime in the following way:
\al{\label{Eq:BMS-like supertranslation}
&
u\to u- f,
&&
r\to r-D^zD_z f,
\nn
&
z\to z+{1\over r}D^z f,
&&
\bar{z}\to \bar{z}+{1\over r}D^{\bar{z}} f.
}
Notice that this is {\it not} the asymptotic symmetry of de Sitter space at a light-like surface with $r \to 1/H$ because the $dudz$ component is different from the de Sitter metric, see also Ref.~\cite{Koga:2001vq}. 
An infinitesimal BMS-like supertranslation changes the metric as follows.
\al{
ds^2&=
\text{(Eq.~\eqref{Eq:static patch})}-rH^2\paren{2\partial^z\partial_z f
-f\partial_u h_{uu} }du^2
-{1\over r}f\partial_u m_B du^2
\nn&+2\paren{\partial^zD_z^2f+{1\over2}f\partial^{z}D_{z}\partial_u U_{z2}-H^2r^2\partial_zf-{1\over2}H^2r^2f\partial_u U_{z2}}dudz
\nn&+2\paren{\partial^{\bar{z}}D_{\bar{z}}^2f+{1\over2}f\partial^{\bar{z}}D_{\bar{z}}\partial_u U_{\bar{z}2}-H^2r^2\partial_{\bar{z}}f-{1\over2}H^2r^2f\partial_u U_{\bar{z}2}}dud\bar{z}
\nn&+\paren{2rD_z^2f-rf \partial_u C_{zz}}dz^2+\paren{2rD_{\bar{z}}^2f-rf \partial_u C_{\bar{z}\bar{z}}}d\bar{z}^2
\nn&+\mathcal{O}(H^2,r^{-2})du^2+\mathcal{O}(H^2,r^{-2})dudr+\mathcal{O}(H,r^{-1})dudz+\mathcal{O}(H,r^{-1})dud\bar{z}
\nn&+\mathcal{O}(1)dzd\bar{z}.
}
From which we obtain the transformation law of each perturbation as
\al{\label{Eq:dS transformation law}
m_B&\to m_B-f\partial_u m_B,\nn
h_{uu}&\to h_{uu}+2\partial^z\partial_zf-f\partial_u h_{uu},\nn
U_{z1}&\to U_{z1}+2\partial^zD_z^2f-f\partial_u U_{z1},\nn
U_{z2}&\to U_{z2}-2\partial_{\bar{z}}f-f\partial_uU_{z2},\nn
C_{zz}&\to C_{zz}+2D_z^2 f-f\partial_u C_{zz}.
}
We find that the BMS-like supertranslation preserves the Bondi gauge and the large $r, H^{-1}$ behavior in Eq. (25).
For $H\to0$ limit, Eqs.~\eqref{Eq:BMS-like supertranslation} and \eqref{Eq:dS transformation law} are smoothly connected to the BMS supertranslation in flat spacetime. 
Again the Einstein equation imposes relations among the various parameters.
As in the flat space case, assuming that $H^{-1}\sim r$, we solve the Einstein equation up to $\mathcal{O}(r^{-3})$, $\mathcal{O}(r^{-2})$, $\mathcal{O}(r^{-1})$ for the $ab$, $aA$ and $aA$ components. This yields
\al{&
h_{uu}=
-{1\over2}\paren{\partial^z U_{z2}+\partial^{\bar{z}} U_{\bar{z}2}},
&&
C_{zz}=-D_z U_{z2}
,
&&
U_{z1}=D^z C_{zz}
.
}
See Appendix~\ref{App:solution} for the detail.
As in Minkowski spacetime, at the vacuum $T_{uu}=\partial_u N_z=0$, the general solution of $C_{zz}$ is $C_{zz}=2D_z^2f$ (under the assumption of the global finiteness of the angular momentum aspect), where $f$ is a real function again, which is nothing but the supertranslation parameter.
Therefore, even in the de Sitter case, the change in the distance between the detectors is parametrized by a BMS-like supertranslation.

The $u$ integration of the Hamiltonian constraint,
\al{\label{Eq: Hamiltonian constraint}
\int^{u_f}_{u_i} du\partial_u m_B&=
\int^{u_f}_{u_i} du\paren{{1\over2}\partial_u\paren{\partial^z U_{z1}+\partial^{\bar{z}} U_{\bar{z}1}}-T_{uu}},
\nn T_{uu}&={1\over4}N_{zz} N^{zz}+4\pi \lim_{r\to\infty}r^2T_{uu}^{\text{(matter)}},
}
gives the explicit parametrization of the  memory by the function $f$ in terms of  the energy momentum flux and the Bondi mass.
Explicitly, it is given by
\al{
f_\text{fin}=-\int d^2z' \gamma_{z'\bar{z}'}G(z,\bar{z};z',\bar{z}')\paren{m_B\big|^{u=u_f}_{u=u_i}+\int^{u_f}_{u_i}du T_{uu}}.
}
This implies that, the memory effect in de Sitter space is same as the flat one if the proper distance between the source and the detector is the same at the time of detection.
The result agrees with that of other authors~\cite{Chu:2015yua,Bieri:2015jwa,Kehagias:2016zry,Chu:2016qxp,Tolish:2016ggo}.

The argument of the zero mode is almost the same as that for the flat background.
One difference is that the asymptotic transformation associated with the zero mode function Eq.~\eqref{Eq:zero mode} is not an isometry of de Sitter spacetime.
However, as long as we focus on the memory effect, this is not a problem because $C_{zz}$ does not change by $f_\text{zero}$, as can be found in Eq.~\eqref{Eq:dS transformation law}.

Finally, we comment on the possible effect of the higher order corrections to the perturbations in Eq.~\eqref{Eq:static patch}, namely, $(rH)^{2m}, m\geq2$ corrections to the metric.
At first sight, such corrections need to be included because we have taken $rH$ to be a finite value.
However, as one can see from the above derivation of the memory effect, the important part is the $r^{-2}$ dependence of 
$uu$
component of the Einstein tensor $G_{uu}$, see also Eq.~\eqref{Eq:uu component} in Appendix~\ref{App:solution}.
This is because $T_{uu}^{\text{(matter)}}$ is proportional to $r^{-2}$ from the energy momentum conservation.
As long as we demand a smooth $H\to0$ limit,
%
the metric $g_{\mu\nu}$ and its inverse $g^{\mu\nu}$ cannot have negative powers of $H$.
Thus, one can argue based on dimensional analysis that the contribution of the higher order $rH$ terms to $G_{uu}$ do not have a $r^{-2}$ dependence.
For example, the $r^3H^4$ correction to the $\Delta g_{uu}=-r^3H^4 h_{uu2}$ gives rise to
\al{
\Delta G_{uu}=r^2H^4\p_u h_{uu2}+\mathcal{O}(r^{-3},H^3),
}
whose $r$ dependence is different from $r^{-2}$, and Eq.~\eqref{Eq:uu component} is not modified.
This implies that higher order terms do not contribute to the 
memory effect.
Thus, our parametrization would remain valid even including higher order terms.

\subsection{Poincare patch}\label{Sec:Poincare patch}
Let us now turn to the Poincare patch.
We show a diffeomorphism that corresponds to the memory effect in Poincare patch.
In Poincare patch, Eq.~\eqref{Eq:static patch} becomes\footnote{
We choose the origin of the conformal time so that the $H\to0$ limit can be taken, which is different from the conventional choice.
See also Appendix~\ref{App:patch}.
}
\al{\label{Eq:de Sitter}
ds^2
&=
\paren{1\over(u+r)H-1}^2\bigg(-(1-h_{uu}^{(P)})du^2+\paren{-2+h_{ur}^{(P)}}dudr+2r^2\gamma_{z\bar{z}}dzd\bar{z}
\nn
& h_{zz}^{(P)} dz^2+h_{\bar{z}\bar{z}}^{(P)}d\bar{z}^2+h_{uz}^{(P)}  dudz+ h_{u\bar{z}}^{(P)}dud\bar{z}+{1\over4}\paren{1+|z|^2}^2h_{zz}^{(P)}h_{\bar{z}\bar{z}}^{(P)}dzd\bar{z}
\bigg)
+...
}
where
\al{\label{Eq:memory Poincare}
h_{uu}^{(P)}&=
-\frac{(H (r+u)-1)}{4 r (H u-1)^2}\bigg\{2 \left(H^2 r^2
   (D^zU_{\text{z2}}+D^{\bar{z}}U_{\bar{z}2})+4 m_B (H(r+u)-1)^2\right)\nn
   &+H D^zU_{\bar{z}2} D^{\bar{z}}U_{z2} (H
   u-1) (H (r+u)-1)\bigg\}
   ,
   \nn
h_{ur}^{(P)}&=   
\left(H\left(r+u\right)-1\right){}^2\frac{D^zU_{\bar{z}2} D^{\bar{z}}U_{z2} }{4 r^2}
   ,
   \nn
h_{uz}^{(P)}&=
(H (r+u)-1)^2\frac{U_{\text{z2}} \left(\frac{H^2r^2}{(H(r+u)-1)^2}-1\right)-D^{\bar{z}}D_{\bar{z}}U_{z2}}{1-
   H u}
   ,
   \nn
h_{zz}^{(P)}&=   
 (H (r+u)-1)r D_zU_{z2}.
}
Here $u=\eta-r$, where $\eta$ is the conformal time and $r$ is the radial direction of the comoving coordinate.
In the vacuum $D_zU_{z2}$ satisfies $D_zU_{z2}=-2D_z^2f$.
Now we consider two detectors 
separated by
 $\delta z$. 
Then, the memory is encoded in the change of the $h_{zz}^{(P)}$, namely, 
\al{
L&\simeq\paren{{1\over(u+r)H-1}}{2r\over1+|z|^2}|\delta z|,\nn
\Delta L&\simeq\paren{1\over(u+r)H-1}^2{1\over2L}\Delta h_{zz}^{(P)}\paren{\delta z}^2+c.c.
={r\over\paren{(u+r)H-1}L} D_z^2f_\text{fin}\paren{\delta z}^2+c.c.
}

In the previous section we have seen that the change of the metric is parametrized by a BMS-like supertranslation.
To derive a transformation corresponding $\Delta h_{zz}^{(P)}$ in the Poincare patch, let us consider the diffeomorphism induced by the coordinate transformation $u\to u+\epsilon_u, r\to r+\epsilon_r, z\to z+\epsilon_z$ and $\bar{z}\to \bar{z}+\epsilon_{\bar{z}}$.
The general infinitesimal diffeomorphism which preserves the Bondi gauge is
\al{\label{Eq:epsilon1}
&
\partial_r\epsilon_u=0,
&&
\epsilon_z=
A_z(u,z,\bar{z})-\frac{ (|z|^2+1)^2}{2 r}\partial _{\bar{z}}\epsilon _u,
&&
\epsilon_{\bar{z}}=
A_{\bar{z}}-\frac{ (|z|^2+1)^2}{2 r}\partial _{z}\epsilon _u,
}
\al{
\label{Eq:epsilon2}
\epsilon_r&=
{1-H(r+u)\over 1-Hu}{\paren{1+|z|^2}^2\over2}\partial_{\bar{z}}\partial_z\epsilon_u
-
{rH\over 1-Hu}\epsilon_u
\nn
&
+
\frac{r(1-H (r+u))}{2\left(1+|z|^2\right) (H u-1)}
\bigg[-2  \paren{ z A_{\bar{z}}+A_z \bar{z} }
+\left(1+|z|^2\right)  \left( \partial
   _zA_z+\partial _{\bar{z}}A_{\bar{z}}\right)
   \bigg].
}
This induces the following transformation of the metric,
\al{
\delta g_{zz}&=
\frac{4 r^2 \partial _z\epsilon
   _{\bar{z}}}{\left(|z|^2+1\right)^2 (H(r+u)-1)^2}.
}   
The solution of the condition $\paren{(u+r)H-1}^{2}\delta g_{zz}=\Delta h^{(P)}_{zz}$ can be obtained by solving\footnote{We thank Chong-Sun Chu for pointing out the mistake in the previous version where we concluded that there is no solution.}
\al{&
D_z^2\epsilon_u={1-Hu\over 2}D_z U_{z2},
&&
\p_z A_{\bar{z}}={H\over2}D^{\bar{z}}U_{z2}.
}
The solution is
\al{&
\epsilon_u=-(1-Hu)f+g,
&&
A_z={H\over4}(1+|z|^2)^2U_{\bar{z}2}+B_z(u,z),
}
where $g$ and $B_z$ are functions which satisfy $D_z^2g=D_{\bar{z}}^2g=0$ and $\p_{\bar{z}}B_z=0$, respectively.
This is the diffeomorphism corresponding to the given memory effect~\eqref{Eq:memory Poincare} in the Poincare patch.
Unlike the case of the flat spacetime, Eq.~\eqref{Eq:az condition}, this transformation contains $\bar{z}$ dependence in $A_z$.
Such breaking of holomorphic structure in superrotation-like parameter $A_z$ reflects the breaking of flat spacetime isometry in de Sitter spacetime, which can be noticed by the fact that $\bar{z}$ dependence is controlled by $H$.

\section{Summary}\label{Sec:Summary}
In this paper, we have considered the relation between memory effect and diffeomorphism in de Sitter space.
As we have seen, a BMS-like supertranslation is useful to parametrize the vacua even in de Sitter spacetime, and therefore the 
intimate relation between memory and BMS-like supertranslation holds also for an accelerating universe.

Our result might have interesting implications to the physics of black hole in de Sitter spacetime.
Our findings suggest that supertranslation(rotation) hair may still be useful to describe the black hole even with the effect of cosmic expansion taken into account.

As for future directions, a few comments are in order.
A natural extension of our work is to look for similar relations in a general FLRW universe.
The decelerating universe case was carried out in Ref.~\cite{Kehagias:2016zry}. It would be interesting to find transformations that parametrize the memory effect in a general setting.
It is often difficult to define the memory effect in curved spacetime because the distinction between memory and tidal force is not easy to make.
Nonetheless, the parametrization of the memory effect by a BMS-like translation may still be useful 
because the effect of tidal force is naturally excluded.
It would be interesting to pursue further this direction.
In this paper, we focus on the memory effect induced by a null messenger particle.
It would certainly be of interest to investigate similarly the memory effect due to massive messengers~\cite{Bieri:2013hqa}. We leave these questions to future study.

\subsection*{Acknowledgement}
We thank Chong-Sun Chu, Yi-Zen Chu and Diego Regalado for useful discussion on the memory effect.
Y.H. and M.S. thank the String Theory and Theoretical Cosmology research group, Department of Physics at the
University of Wisconsin-Madison and the Institute for Advanced Study of the Hong Kong University of Science and Technology for hospitality during their visits. 
The work of Y.H. is supported by the Grant-in-Aid for JSPS Fellows No.16J06151.
M.S. is supported by IBS under the project code, IBS-R018-D1.
The work of G.S. is supported in part by the DOE grant DE-FG- 02-95ER40896, and the Kellett Award of the University of Wisconsin.

\appendix
\section{Solution in the static patch of de Sitter}\label{App:solution}
In this appendix, we present the detailed calculation in the de Sitter static patch.
The starting line element is
\al{\label{Eq:static patch2}
ds^2&=
(-1+r^2H^2)du^2-2dudr+2r^2\gamma_{z\bar{z}}dzd\bar{z}
\nn&+\br{{m_B(u,z,\bar{z})\over r}-rH^2h_{uu}(u,z,\bar{z})}du^2+rC_{zz}(u,z,\bar{z})dz^2+rC_{\bar{z}\bar{z}}(u,z,\bar{z})d\bar{z}^2
\nn&+\br{U_{z1}(u,z,\bar{z})+H^2r^2U_{z2}(u,z,\bar{z})}dudz
\nn&+\br{U_{\bar{z}1}(u,z,\bar{z})+H^2r^2U_{\bar{z}2}(u,z,\bar{z})}dud\bar{z}+{1\over r^2}\paren{D_{ur1}+H^2r^2D_{ur2}}dudr
\nn&+{1\over r}\paren{{4\over3}N_z+{4\over3}u\partial_z m_B-{1\over4}\partial_z(C_{zz}C^{zz})}dudz+{\paren{1+|z|^2}^2\over2}C_{zz}C_{\bar{z}\bar{z}}dzd\bar{z}+....
}
Each function depends on $(u,z,\bar{z})$, and does not depend on $r$ and $H$.

In the $H\to0$ limit, the solution has to be same as that in flat spacetime.
Therefore, it is found that
\al{&
U_{z1}=D^z C_{zz},
&&
D_{ur1}={\paren{1+|z|^2}^4\over16}C_{zz}C_{\bar{z}\bar{z}}+h(z,\bar{z}).
}
Next, let us solve the Einstein equation in a large $r, H^{-1}$ expansion.
We expand the Einstein tensor in powers of $1/r$ and $H$, while keeping the ratio $Hr$ at a finite value. 
As mentioned in Sec.~\ref{Sec:BMS}, we solve the 
$ab$, $aA$ and $aA$ components of the
Einstein equation up to $\mathcal{O}(r^{-3})$, $\mathcal{O}(r^{-2})$ and $\mathcal{O}(r^{-1})$, respectively.
The $(u,r)$ and $(z,\bar{z})$ components of the Einstein equation give
\al{
-{1\over2}{(rH)^2\over r^3}\br{4h_{uu}+(1+|z|^2)^2\paren{\partial_{\bar{z}}U_{z2}+\partial_{z}U_{\bar{z}2}}}&=0,\nn
{(rH)^2\over2r(1+|z|^2)^2}\br{4h_{uu}+(1+|z|^2)^2\paren{\partial_{\bar{z}}U_{z2}+\partial_{z}U_{\bar{z}2}}}&=0,
}
respectively.
Then, we obtain
\al{
h_{uu}=
-{1\over2}\paren{\partial^z U_{z2}+\partial^{\bar{z}} U_{\bar{z}2}}.
}
Finally, the $(z,z)$ component gives the relation
\al{
-{(rH)^2\over r}\paren{C_{zz}+D_zU_{z2}}=0,
}
from which we get $C_{zz}=-D_zU_{z2}$.
At this stage, we can see that the Einstein equation is satisfied except for the $(u,u)$ and $(u,z)$ components.
At $\mathcal{O}(r^{-2})$, the $(u,u)$ component is
\al{\label{Eq:uu component}
{1\over 2r^2}\br{-4\partial_u m_B-N_{zz}N^{zz}+\partial_u\paren{\partial^zU_{z1}+\partial^{\bar{z}}U_{\bar{z}1}}}
=8\pi T_{uu}^\text{(matter)},
}
which plays a crucial role in parametrizing the memory effect.
The remaining components, the $\mathcal{O}(r^{-3})$ term in the $(u,u)$ component and the $\mathcal{O}(r^{-2})$ term in the $(u,z)$ component, just determine the $u$ dependence of  the $\mathcal{O}(r^{-2})$ term in $g_{uu}$ and the $\mathcal{O}(r^{-1})$ term in $g_{uz}$, respectively.

In the beginning of our calculation Eq.~\eqref{Eq:static patch2}, we have assumed the absence of odd powers of $H$ in the perturbation terms.
This may be justified because the de Sitter Schwarzschild background and the source term $T_{uu}$ do not break the symmetry $H\to-H$.

\section{Patches in de Sitter space}\label{App:patch}
Here we collect some useful formulas describing describing the different patches of de Sitter spacetime.
See, e.g., Ref.~\cite{Spradlin:2001pw} for a review.
The de Sitter space is defined as the hyperboloid
\al{
-X_0^2+X_1^2+X_2^2+X_3^2+X_4^2=H^{-2},
}
which is embedded in $5$ dimensional flat spacetime $X^{0-4}$.
\begin{itemize}
\item
Global patch

In the global patch, the $5$ dimensional coordinates are parametrized as
\al{&
X_0=H^{-1}\sinh(\tau H),
&&
X_1=H^{-1}\cosh(\tau H)\cos\theta_1,\nn
&
X_2=H^{-1}\cosh(\tau H)\sin\theta_1\cos\theta_2,
&&
X_3=H^{-1}\cosh(\tau H)\sin\theta_1\sin\theta_2\cos\theta_3,\nn
&
X_4=H^{-1}\cosh(\tau H)\sin\theta_1\sin\theta_2\sin\theta_3.
}
The metric is
\al{
ds^2=-d\tau^2+H^{-2}\cosh^2(\tau H)d\Omega_3,
}
where $d\Omega_3=d\theta_1^2+\sin^2\theta_1\paren{d\theta_2^2+\sin^2\theta_2d\theta_3^2}$.

\item
Conformal patch

Sometimes it is convenient to use $T=\arcsin\paren{\tanh(\tau H)}$.
In this case, 
\al{
ds^2={1\over H^2\cos^2T}\paren{-dT^2+d\Omega_3},
}
which is used to write the Penrose diagram of de Sitter space.

\item
Static patch

In the static patch, the $5$ dimensional coordinates are parametrized as
\al{&
X_0=\sqrt{H^{-2}-r^2}\sinh (t H),
&&
X_1=r \cos\theta_1,
&&
X_2=r \sin\theta_1\cos\theta_2,\nn
&
X_3=r  \sin\theta_1\sin\theta_2,
&&
X_4=\sqrt{H^{-2}-r^2}\cosh (t H).
}
The metric is
\al{
ds^2=-(1-r^2H^2)dt^2+{dr^2\over1-r^2H^2}+r^2d\Omega_2.
}
\item
Poincare patch

In Poincare patch, the $5$ dimensional coordinates are parametrized as
\al{&
X_0={H^{-2}-\eta'^2+x_1^2+x_2^2+x_3^2\over -2\eta'},
&&
X_1=H^{-1}{x_1\over -\eta'},\nn
&
X_2=H^{-1}{x_2\over -\eta'},
&&
X_3=H^{-1}{x_3\over -\eta'},\nn
&
X_4={H^{-2}+\eta'^2+x_1^2+x_2^2+x_3^2\over -2\eta'}.
}
The metric is
\al{
ds^2=H^{-2}{1\over\eta'^2}\paren{-d\eta'^2+dx_1^2+dx_2^2+dx_3^2}.
}
In this paper, we have introduced $\eta:=\eta'+H^{-1}$, and the line element is
\al{
ds^2={1\over\paren{\eta H-1}^2}\paren{-d\eta^2+dx_1^2+dx_2^2+dx_3^2},
}
which is used in Sec.~\ref{Sec:Poincare patch}.
\end{itemize}

\bibliographystyle{TitleAndArxiv}
\bibliography{Bibliography}


\end{document}